\documentclass[11pt]{article}
\usepackage{arxiv}

\usepackage{amsmath}  
\usepackage{graphicx}
\usepackage{filecontents}
\usepackage{multirow}
\usepackage{xcolor}
\usepackage{soul}
\usepackage{textcomp}
\title{Ignatyuk damping factor: A semiclassical formula }

\author{
  Nishchal R. Dwivedi$^{1,2}$, Saniya Monga$^{3}$, Harjeet Kaur$^{3}$ and~Sudhir R. Jain$^{1,4,5}$ \\
  $^1$Nuclear Physics Division, Bhabha Atomic Research Centre, Trombay, Mumbai 400 085, India\\
  $^2$Department of Physics, University of Mumbai, Vidyanagari Campus, Mumbai 400 098, India\\
  $^3$Department of Physics, Guru Nanak Dev University, Amritsar 143 005, India\\
  $^4$Homi Bhabha National Institute, Training School Complex, Anushakti Nagar, Mumbai 400 094, India \\
  $^5$UM-DAE Centre for Excellence in Basic Sciences, University of Mumbai, Mumbai 400 098, India\\
}

\begin{document}
\maketitle
\begin{abstract}

Data on nuclear level densities extracted from transmission data or gamma energy spectrum store the basic statistical information about nuclei at various temperatures. Generally this extracted data goes through model fitting using computer codes like CASCADE. However, recently established semiclassical methods involving no adjustable parameters to determine the  level density parameter for magic and semi-magic nuclei give a good agreement with the experimental values. One of the popular ways to paramaterize the level density parameter which includes the shell effects and its damping was given by Ignatyuk. This damping factor is usually fitted from the experimental data on nuclear level density and it comes around 0.05 $MeV^{-1}$. In this work we calculate the Ignatyuk damping factor for various nuclei using semiclassical methods.
\end{abstract}


\maketitle
\section{Introduction}
Use of statistical ideas, for understanding nuclei comprising of closely spaced energy levels in their excited states, was proposed by Bohr \cite{bohr}. This is subsumed in the usage of level density for understanding the distribution of neutron resonances, evaluating reaction cross sections and other experimental observables  \cite{lang}. The study of level density has been of interest even till the recent times \cite{zelevinsky}.  Most of the computer codes which are utilized to evaluate the level density of a nucleus are based on back-shifted Fermi gas (BSFG) formula \cite{bethe} and constant temperature (CT) formula \cite{eric}.
The free  parameters involved in these formulae can be experimentally obtained by fitting known energy levels of complete level schemes at low excitation energies together with neutron resonances at the neutron binding energy \cite{al,Audi,Gutt}. Gilbert and Cameron \cite{gil} proposed a formula for nuclear level density, composed of four parameters, which combines the BSFG formula at high excitation energies with a CT formula for lower energies. By fitting the four constants in both regions, experimental data may be well-reproduced \cite{kawano}.

However, all these models may not explain well the dependence of the level density on quantities of significant interest like temperature, deformations \cite{roy1} and angular momentum \cite{roy2}. Phenomenological relations are used to describe the thermal damping of the shell effects with increasing excitation energies  \cite{Ignatyuk, kataria, vonach} by exploiting microscopic calculations for level density parameter to evaluate nuclear level density. 
One such widely used expression for fitting the excitation energy ($E^{*}$) dependent level density parameter, $a(E^{*})$, by incorporating the role of shell effects was given by Ignatyuk \cite{Ignatyuk} as,

\begin{eqnarray}
a(E^{*})=\tilde{a}\left[1+\frac{\delta W}{E^{*}}\{1-\exp(-\gamma E^{*})\}\right]
\label{ign}
\end{eqnarray}
where $\tilde{a}=A/n$ $MeV^{-1}$ is the asymptotic value of nuclear level density parameter. $\gamma$ is called the damping coefficient and $\delta W$ denotes the shell correction energy which is given as the difference between the experimental binding energy of a nucleus and the binding energy calculated from the liquid drop model \cite{nndc}. The damping coefficient $\gamma$, has no known exact expression and is often employed as a fitting parameter. It has the typical value between 0.04 to 0.07 MeV$^{-1}$ (see for example, \cite{rout}).

Semiclassical methods have been very useful in understanding complex systems like deformed nuclei \cite{Bohr-Mottelson2,jj}. These have been gainfully employed to obtain analytic expression for the ground states of deuteron \cite{deuteron} and triton \cite{triton}, in understanding shell effects in the light of chaos \cite{zel}, nuclear dissipation \cite{jp,npa}, nuclear cluster energies in astrophysical context \cite{gulminelli} and even studying statistical properties of the nuclei \cite{suraud}. Recent work \cite{kaur} on the melting of shell effects in  magic and semi-magic nuclei with excitation energy, based on semiclassical trace formulae derived by Brack and Jain \cite{brack-jain}, has resulted in level density parameter within 10\%-15\% of the experimental values and BSFG model calculations with \textit{no} adjustable parameters. 

Here, we continue this line of work to evaluate the damping coefficient appearing in Ignatyuk prescription (\ref{ign}) which is employed extensively to quantify the disappearance of shell effects at higher excitation energies. For $Ni$-, $Sn$- and $Pb$- isotopes, we also study the energy average of the damping coefficient from $0-10$ $MeV$ of excitation energy  and its variation with the $A/n$ model in a phenomenological way. In Table \ref{tab}, for $\tilde{a}=A/8$ $MeV^{-1}$, the calculated average values of $\gamma$ are given. The variation of damping coefficient with excitation energy $E^*$ for different $n$ is studied for some $Pb$- isotopes in Fig. \ref{fig:PbE}. We further investigate the asymptotic behaviour of $n$ in the regime where it is known that the shell effects \textit{`melt'} due to high excitation energies. These studies may hint towards shape transitions and how the nuclear potential changes.

\section{Semiclassical Ignatyuk Formula}

The expression (\ref{ign}) has variables of excitation energy $E^{*}$, excitation energy dependent level density parameter $a(E^{*})$, the parameter $n$ used in the asymptotic model of $\tilde{a}=A/n$ $MeV^{-1}$, and the damping coefficient $\gamma$. $E^{*}$ and  $a(E^{*})$, can be determined semiclassically \cite{kaur}. Then for a set value of $n$, we can evaluate damping coefficient $\gamma$.

The level density  parameter $a(T)$ at finite temperature $T$  is given as \cite{Bohr-Mottelson1},

\begin{eqnarray}
a_T=\frac{\pi^2}{6} g_T(\mu)
\label{at}
\end{eqnarray}
where  $g_T(\mu)$ is the finite-temperature single-particle level density of the system at chemical potential $\mu$. For realistic nuclear level density parameter calculations, we exploit the single-particle level density for a spherically symmetric harmonic oscillator potential with spin-orbit interactions at finite temperature \cite{kaur} which may be split as,

\begin{eqnarray}
g_T(E)=\tilde{g_{T}}(E)+\delta g_{T}(E)
\end{eqnarray}
with average part of level density \cite{brack},
\begin{eqnarray}
\tilde{g_{T}}(E)&=&\frac{1+3\kappa^2 \hbar^2 \omega_0^2 }{2\hbar^3 \omega_0^3}\left(E^2+\frac{\pi^2 T^2}{3}\right)\nonumber\\
&-&\frac{\left(1+5\kappa^2 \hbar^2 \omega_0^2 \right)}{8\hbar \omega_0}+ \kappa^3 \hbar \omega_0 T ln[1+e^{E/T}]+... 
\label{levd}
\end{eqnarray}
and, the oscillatory part of the single-particle level density  which incorporates the shell effects in the nuclei is given as \cite{kaur}:
\begin{eqnarray}
\delta g_{T}(E)&=&\frac{1}{(2\hbar \omega_0)^2}\sum_{\pm}\sum_{k=1}^{\infty} \frac{E\pi k T T_{\pm}\sin\left(\frac{k E T_{\pm}}{\hbar}-\frac{k \pi \sigma_{\pm}}{2}\right)}{\hbar(1\mp\kappa \hbar \omega_0)^2\sinh(\pi k T T_{\pm}/\hbar)\sin(2 k \pi/\{1\mp \kappa \hbar \omega_0\})} \nonumber \\
&+&\frac{1}{(2\hbar \omega_0)^2}\sum_{\pm}\sum_{k=1}^{\infty}\frac{\left[(\pi T)^2 k T_{\pm}\coth{\left(\pi k T T_{\pm}/\hbar\right)-\pi T \hbar}\right]\cos\left(\frac{k E T_{\pm}}{\hbar}-\frac{k \pi \sigma_{\pm}}{2}\right)}{\hbar(1\mp\kappa \hbar \omega_0)^2\sinh(\pi k T T_{\pm}/\hbar)\sin(2 k \pi/\{1\mp \kappa \hbar \omega_0\})} \nonumber \\
 &+&\frac{1}{2\hbar \omega_0}\sum_{\pm}\sum_{k=1}^{\infty}\frac{(\mp 1+2 \kappa \hbar \omega_0)\pi k TT_{\pm}/\hbar}{2(1\mp \kappa \hbar \omega_0)^2\sin\left({2 k \pi}/\{1\mp \kappa \hbar \omega_0 \}\right) \sinh \left({\pi k T_{\pm} T}/{\hbar}\right)}\sin \left(\frac{k T_{\pm} E}{\hbar}-\frac{k \pi \sigma_{\pm}}{2}\right)   \nonumber \\
&+&\frac{1}{2\hbar \omega_0}\sum_{\pm}\sum_{k=1}^{\infty}\frac{\cos \left({2 k \pi}/\{1\mp \kappa \hbar \omega_0\}\right)\pi k T T_{\pm}/\hbar  }{(1\mp \kappa \hbar \omega_0)^2\sin^2 \left({2 k \pi}/\{1 \mp \kappa \hbar \omega_0\}\right) \sinh \left(\pi k T_{\pm} T/{\hbar}\right)}\cos \left(\frac{k T_{\pm} E}{\hbar}-\frac{k \pi \sigma_{\pm}}{2}\right) \nonumber \\
&+& \frac{1}{2\hbar \omega_0}\sum_{k=1}^{\infty}\frac{(-1)^{k+1}\pi k T T_{0}/\hbar }{2 \sin^2 {(k \pi \kappa \hbar \omega_0)}  \sinh \left({\pi k T_0 T}/{\hbar}\right)}\cos\left(\frac{k T_{0} E} {\hbar}-\frac{ k \pi \sigma_{0}}{2}\right) \nonumber \\
&+&\sum_{s^{\prime}=1}^{\infty}(-1)^{s^{\prime}}\exp(-s^{\prime}E/T)\times \Delta_4 
\label{spinl}
\end{eqnarray}

where $  \Delta_4$ is given by
\begin{eqnarray}
\Delta_4 &=&\frac{1}{(2\hbar \omega_0)^2}\sum_{\pm}\sum_{k=1}^{\infty}\frac{\left(2\left(\frac{s^{\prime}}{T}\right)^2\frac{k T_{\pm}}{\hbar} \cos \left({k \pi \sigma_{\pm}}/{2}\right)\right)+\frac{s^{\prime}}{T}\left(\left(\frac{s^{\prime}}{T}\right)^2-\left(\frac{kT_{\pm}}{\hbar}\right)^2\right)\sin(k \pi \sigma_{\pm} /2)}{(1\mp \kappa \hbar \omega_0)^2 \sin \left({2 k \pi}/\{1\mp \kappa \hbar \omega_0\}\right)\left(\left(\frac{k T_{\pm}}{\hbar}\right)^2+\left(\frac{s^{\prime}}{T}\right)^2\right)^2}\nonumber \\
&+&\frac{1}{2\hbar \omega_0}\sum_{\pm}\sum_{k=1}^{\infty}\frac{(\mp 1+2 \kappa \hbar \omega_0)}{2(1\mp \kappa \hbar \omega_0)^2}\frac{\left(\frac{s^{\prime}}{T}\right)^2 \sin \left({k \pi \sigma_{\pm}}/{2}\right)-\frac{s^{\prime}kT_{\pm}}{T\hbar}\cos\left(k \pi\sigma_{\pm}/2\right)}{ \sin \left({2 k \pi}/\{1\mp \kappa \hbar \omega_0\} \right) \left(\left(\frac{k T_{\pm}}{\hbar}\right)^2+\left(\frac{s^{\prime}}{T}\right)^2\right)}\nonumber \\
&+&\frac{1}{2\hbar \omega_0}\sum_{\pm}\sum_{k=1}^{\infty}\frac{\cos \left({2 k \pi}/\{1\mp  \kappa \hbar \omega_0 \}\right)}{\sin^2 \left({2 k \pi}/\{1\mp \kappa \hbar \omega_0\}\right)}\frac{\left(\frac{s^{\prime}}{T}\right)^2 \cos \left({k \pi \sigma_{\pm}}/{2}\right)-\left(\frac{s^{\prime}kT_{\pm}}{T\hbar}\right)\sin(k \pi \sigma_\pm /2)}{(1\mp \kappa \hbar \omega_0)^2\left(\left(\frac{k T_{\pm}}{\hbar}\right)^2+\left(\frac{s^{\prime}}{T}\right)^2\right)} \nonumber \\
&+&\frac{1}{2\hbar \omega_0}\sum_{k=1}^{\infty}(-1)^{k+1}\frac{\left(\frac{s^{\prime}}{T}\right)^2 \cos \left({k \pi \sigma_0}/{2}\right)-\left(\frac{s^{\prime} k T_0}{T\hbar}\right)\sin\left(k \pi \sigma_0/2\right)}{ 2\sin^2 \left(k \pi \kappa \hbar \omega_0 \right) \left(\left(\frac{s^{\prime}}{T}\right)^2+\left(\frac{k T_0}{\hbar}\right)^2\right)}. \nonumber
  \end{eqnarray}
$\kappa$ denotes the strength of spin-orbit interactions and it is measured in units of $(\hbar\omega_0)^{-1}$.

For nuclei with equal number of neutrons $(N)$ and protons $(Z)$ ($N=Z=A/2$), chemical potential $\mu$ is fixed as:
\begin{eqnarray}
\frac{A}{2}=\int_0^{\mu}g_T(E)dE
\end{eqnarray}
where $A$ is the mass number of nucleus. For nuclei with $N \neq Z$, the chemical potentials $\mu^{(N,Z)}$'s are fixed by neutron and proton number respectively, by employing 
\begin{eqnarray}
N = \int_0^{\mu^{N}} g_T(E) dE. \nonumber \\
Z = \int_0^{\mu^{Z}} g_T(E) dE.
\label{mu}
\end{eqnarray}
Once we fix the chemical potentials, the total level density parameter $a_T$ obtained by utilizing the total finite-temperature level density $g_T(E)$ \cite{kaur} is given as:
\begin{eqnarray}
a_T=\frac{\pi^2}{6} \left[g_T(\mu^N)+g_T(\mu^Z)\right]. 
\label{ldp}
\end{eqnarray}

The excitation energy $E^{*}$ is given as the difference between the internal energies at finite temperature and at zero temperature as,
\begin{eqnarray}
E_{N,Z}^{*}&=&E_{N,Z}(T)-E_{N,Z}(T=0)\nonumber \\
 &=&\int_0^{\infty}Eg(E)\left[1+\exp{\left(\frac{E-\mu^{N,Z}}{T}\right)}\right]^{-1} dE\nonumber\\
 &-&\int_0^{\epsilon_{F}^{N,Z}}E g(E) dE
 \label{exc}
\end{eqnarray}
where $g(E)=\tilde{g}(E)+\delta g(E)$, is the zero temperature single-particle level density \cite{amann} given as,
\begin{eqnarray}
\tilde {g}(E)&=&\frac{E^2}{2\hbar^3 \omega_0^3}\Biggl[1+3 \kappa^2 \hbar^2 \omega_{0}^2 \Biggr]-\frac{1}{8\hbar\omega_0}\left[1+5 \kappa^2\hbar^2 \omega_{0}^2  \right]\nonumber\\
&+&{E}\kappa^3 {\hbar \omega_0}+O(\hbar^4 \kappa^4)+....
\label{zero}
\end{eqnarray}
 and the oscillatory part of the level density for such a system is given by \cite{amann}:  
\begin{eqnarray}
\delta g(E)&=&\frac{E}{2(\hbar\omega_0)^2}\sum_{\pm}\sum_{k=1}^{\infty}\frac{\sin\left(\frac{kET_{\pm}}{\hbar}-\frac{k\pi\sigma_{\pm}}{2}\right)}{(1\mp \kappa \hbar\omega_0)^2
\sin\left(2k \pi/\{1\mp\kappa \hbar\omega_0\}\right)}\nonumber \\
&+&\frac{1}{2\hbar\omega_0}\sum_{\pm}\sum_{k=1}^{\infty}\frac{(\mp 1+2\kappa\hbar\omega_0)\sin\left(\frac{kET_{\pm}}{\hbar}-\frac{k\pi\sigma_{\pm}}{2}\right)}{2(1\mp \kappa \hbar\omega_0)^2
\sin\left(2k \pi/\{1\mp \kappa \hbar\omega_0\}\right)}\nonumber \\
&+&\frac{1}{2\hbar\omega_0}\sum_{\pm}\sum_{k=1}^{\infty}\frac{\cos\left(\frac{2k\pi}{1\mp\kappa\hbar\omega_0}\right)\cos\left(\frac{kET_{\pm}}{\hbar}-\frac{k\pi\sigma_{\pm}}{2}\right)}{(1\mp \kappa \hbar\omega_0)^2
\sin^2\left(2k \pi/\{1\mp\kappa \hbar\omega_0\}\right)}\nonumber \\
&+&\frac{1}{2\hbar\omega_0}\sum_{k=1}^{\infty}\frac{(-1)^{k+1}}{2\sin^2(k\pi\kappa\hbar\omega_0)}\cos\left(\frac{kET_0}{\hbar}-\frac{k\pi\sigma_0}{2}\right)\nonumber\\
\nonumber
\end{eqnarray}
with three time periods, 
\begin{eqnarray}
T_{\pm}=\frac{2\pi}{\omega_0(1\mp \kappa\hbar\omega_0)}, \qquad T_0=\frac{2\pi}{\omega_0} \nonumber
\end{eqnarray}
and non-integer Maslov indices as
\begin{eqnarray}
\sigma_{\pm}=\frac{\mp 2}{1\mp\kappa\hbar\omega_0}, \qquad \sigma_0=-4\kappa\hbar\omega_0. \label{zer}
\end{eqnarray}
$T_0$  is the time period for the classical orbits of the unperturbed Hamiltonian and the periodic orbits of the perturbed system are described by the shifted time periods $T_{\pm}$. $k$ is the number of repetitions of the periodic orbits over which the sum has to be calculated.

From equation (\ref{exc}), we can find the excitation energy as a function of temperature and consequently, we can have excitation energy dependent level density parameter $a(E^{*})$. These excitation energies $E^{*}$ are used in \eqref{gam} along with corresponding level density parameters $a(E^{*})$ to find $\gamma$  at various excitation energies. The method of finding $E^*$ and $a(E^*)$  using the semiclassical trace formula \cite{kaur} for magic and semi-magic nuclei having no adjustable parameters, leading to temperature dependent level density parameter to be exact.

We can invert the relation (\ref{ign}) to evaluate the Ignatyuk damping coefficient $\gamma$ as following:
\begin{eqnarray}
\gamma=-\frac{1}{E^{*}}\log_e\left[1+\frac{E^{*}}{\delta W}\left(1-\frac{a(E^{*})}{\tilde{a}}\right)\right].
\label{gam}
\end{eqnarray}

\section{Calculations and results}
To determine the damping coefficients for $Ni$-, $Sn$- and $Pb$- isotopes according to this formalism,  we have used the values of spin-orbit interaction strength parameters, $\kappa_N$ and $\kappa_Z$ as given in \cite{rag,kaur}. 

\begin{table}[ht!]
    \centering
    \begin{tabular}{c c c c c c }
 \hline
Nucleus & $  \kappa_{N,Z} (\hbar\omega_{N,Z})^{-1}$ & $\delta W$ $(MeV)$  & $\gamma$ $(MeV)^{-1}$ \\
 \hline
 \hline
  $^{48}Ni$ &-0.080, -0.075 & -8.20 $\pm$0.48 &  0.022 $\pm$ 0.028   \\
 $^{58}Ni$ & -0.090, -0.075  & -1.70497 $\pm$ 0.00034 & 0.03896 $\pm$ 0.00004 \\  
 $^{112}Sn$ &-0.070, -0.060    & 0.25838 $\pm$ 0.00033 & 0.013071$\pm$ 0.00092 \\
 $^{114}Sn$ &-0.070, -0.060   & 0.78511 & 0.03386 \\
 $^{116}Sn$ &-0.070, -0.060    & 0.97313 $\pm$ 0.00012 & 0.04858$\pm$ 0.00008 \\
 $^{182}Pb$ & -0.134, -0.151 & -1.81636 $\pm$ 0.01274 & 0.07194 $\pm$ 0.00640 \\
 $^{204}Pb$ & -0.134, -0.151  & -7.22956$\pm$ 0.00122 & 0.019396 $\pm$ 0.00008 \\
 $^{206}Pb$ & -0.134, -0.151  & -8.80197 $\pm$ 0.00124 & 0.02284 $\pm$ 0.00007\\
 $^{208}Pb$ & -0.134, -0.151 & -10.255 $\pm$ 0.00125 &  0.027179 $\pm$ 0.00015\\
 $^{210}Pb$ & -0.134, -0.151 &-7.30008 $\pm$ 0.00147 &0.027511 $\pm$ 0.00013\\
 \hline
 \hline
\end{tabular}
    \caption{By using the shown spin-orbit interaction strength parameters and shell correction energies, the average $\gamma$ values obtained for $Ni$, $Sn$ and $Pb$ isotopes for 0-10 MeV excitation energy taking $\tilde{a}=A/8$ $MeV^{-1}$ are given in this Table.}
    \label{tab}
\end{table}

The energy unit $\hbar \omega_0$ is given as following \cite{rag}:
\begin{eqnarray}
\hbar \omega_0 (N,Z) = \frac{41}{A^{1/3}}\left(1\pm \frac{N-Z}{A}\right) {\mathrm{MeV}}.
\end{eqnarray}
The shell correction energy, $\delta W$, corresponding to the difference between experimental and liquid drop model binding energy is taken from \cite{nndc}. The parameter, $k$, describes the repetitions over periodic orbits and we have set its value by assuming $a_T=\tilde{a}$ at temperature $T=5 $ MeV \cite{kaur}.

\begin{figure}[ht!]
\centering
\includegraphics{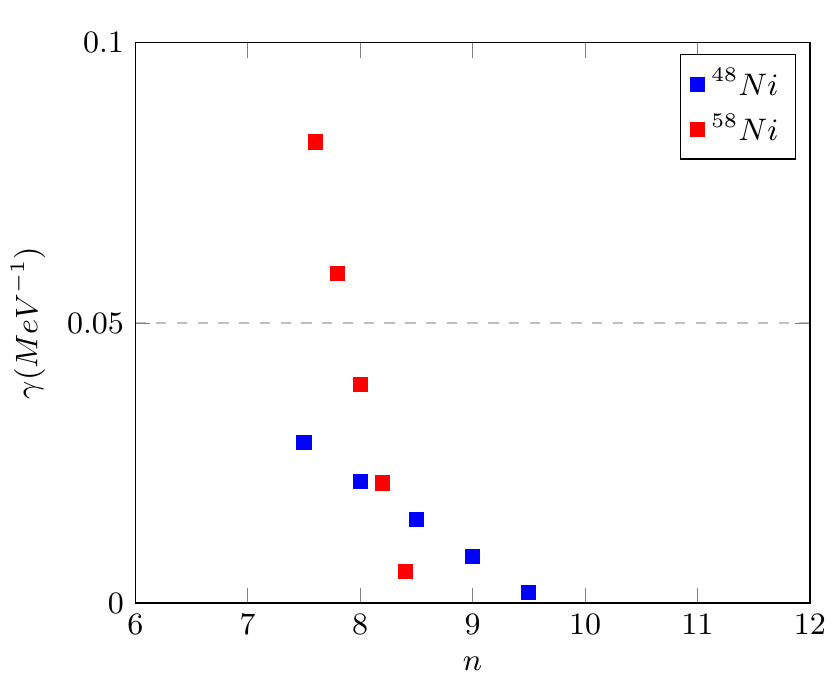}
\caption{The variation of $\gamma$ with $n=A/\tilde{a}$ for $Ni$- isotopes. Values of $\gamma$ decrease with the increase in $n$ and it approach to  zero for  $^{48}Ni$ and $^{58}Ni$, at $n=8.4$ and $n=9.5$, respectively, symbolizing a nuclear model with no shell effects, or when shell effects have completely vanished. $^{48}Ni$ also attains spherical symmetry $(n=10)$ when shell effects are assumed to be completely washed off.}
\label{fig:Ni}
\end{figure}

\begin{figure}[ht!]
\centering

\includegraphics{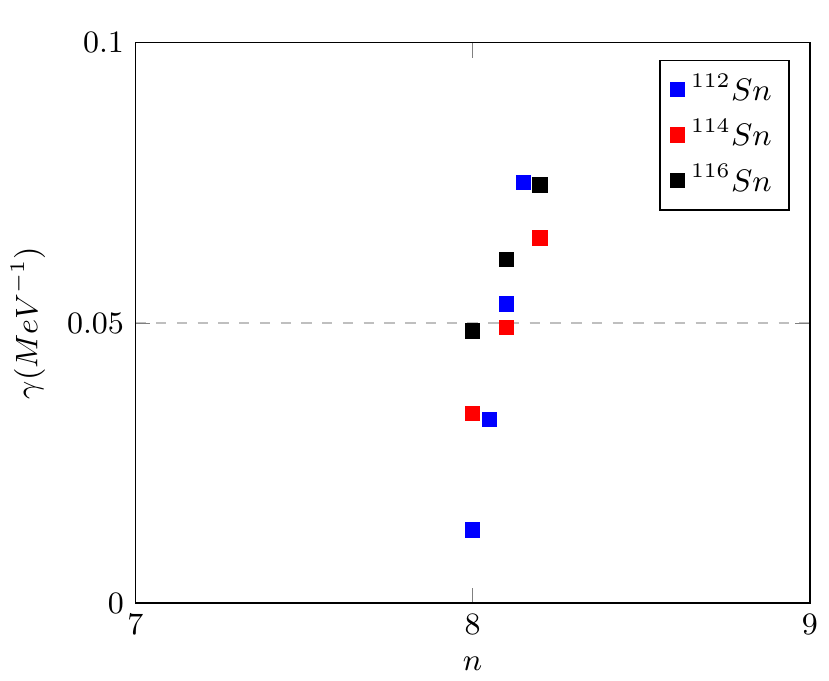}
\caption{The variation of values of $\gamma$ with $n=A/\tilde{a}$ for $Sn$-isotopes. The values of the damping factor are quite sensitive for $^{112, 114, 116}Sn$ between $n=8-9$, due to their $\delta W$ values being closer to zero as compared to that of other considered nuclei, which exhibits a smaller slope.}
\label{fig:Sn}
\end{figure}

\begin{figure}[ht!]
\centering
\includegraphics{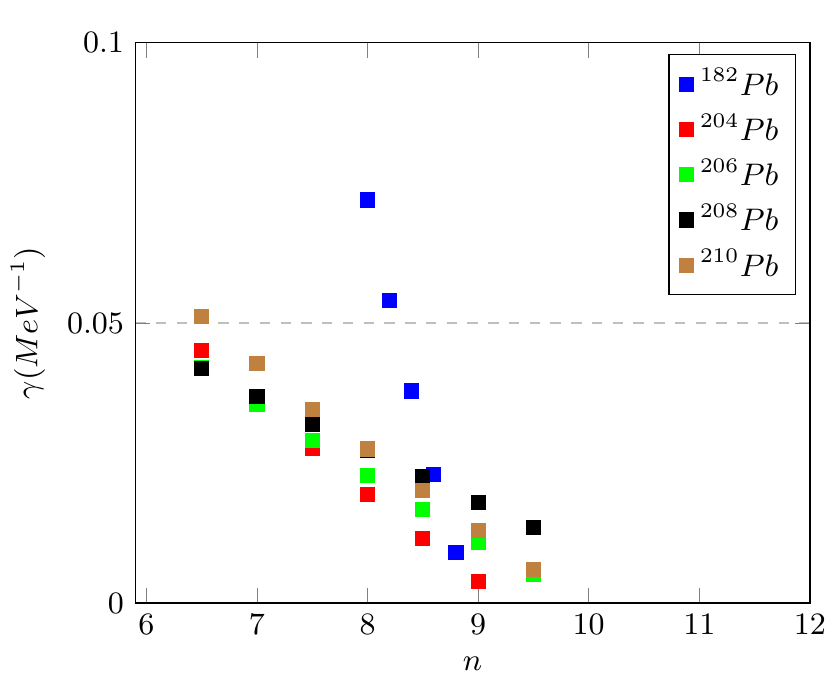}
\caption{The variation of values of $\gamma$ with $n=A/\tilde{a}$ for $Pb$-isotopes. At $n=6$, the values of $\gamma$ are maximum (8 for $^{182}Pb$), and it tends to zero as $n$ goes towards 10. Since $n=10$ corresponds to harmonic oscillator potential, we can say that as the system approaches a spherical symmetry, the shell contribution reduce for $Pb$- isotopes.}
\label{fig:Pb}
\end{figure}


 




The excitation energy $E^{*}$ as a function of temperature $T$ is evaluated utilizing expression (\ref{exc}) after evaluation of chemical potentials corresponding to neutrons and protons according to (\ref{mu}). Also, the level density parameter is determined using (\ref{ldp}) for $N\neq Z$.

In \eqref{gam}, the excitation energy, $E^*$, is a continuous variable and the only free parameter is in the asymptotic level density parameter $\tilde{a} = A/n$ MeV$^{-1}$, which is usually taken between $A/7$ to $A/10$ MeV$^{-1}$. We calculate $\gamma$ for ten nuclei involving $Ni$-, $Sn$- and $Pb$- isotopes for various excitation energies between $0  - 10 ~ $MeV. The average of these values for different $n$'s are plotted in Figs. \ref{fig:Ni}, \ref{fig:Sn} and  \ref{fig:Pb}. The typical values of damping coefficient using $n=8$ are given in Table \ref{tab}. These values have been studied experimentally for the $Pb$- isotopes, and our evaluated values lie close to this experimentally accepted range \cite{rout}.

A nuclear potential corresponding to an isotropic harmonic oscillator potential corresponds to  $\tilde{a}=A/10$, which has an analogy to spherical nuclear shapes \cite{akjain}. The semiclassical study in \cite{kaur}, for harmonic oscillator potential with spin-orbit interactions also asserts this where the theoretically value of $n$ comes out to be 10 for over 90 magic/semi-magic nuclei.

In $Ni$ isotopes  (Fig.\ref{fig:Ni}), we see a decreasing trend in $\gamma$ as the $n$, changes. The damping parameter in $^{48}Ni$ and $^{58}Ni$ approach to zero at around $n=8.4$ and $n=9.5$, respectively. This hints towards agreement with approaching the spherical shape as the shell effects contribution decreases.

\begin{figure}[ht!]
\centering
\includegraphics{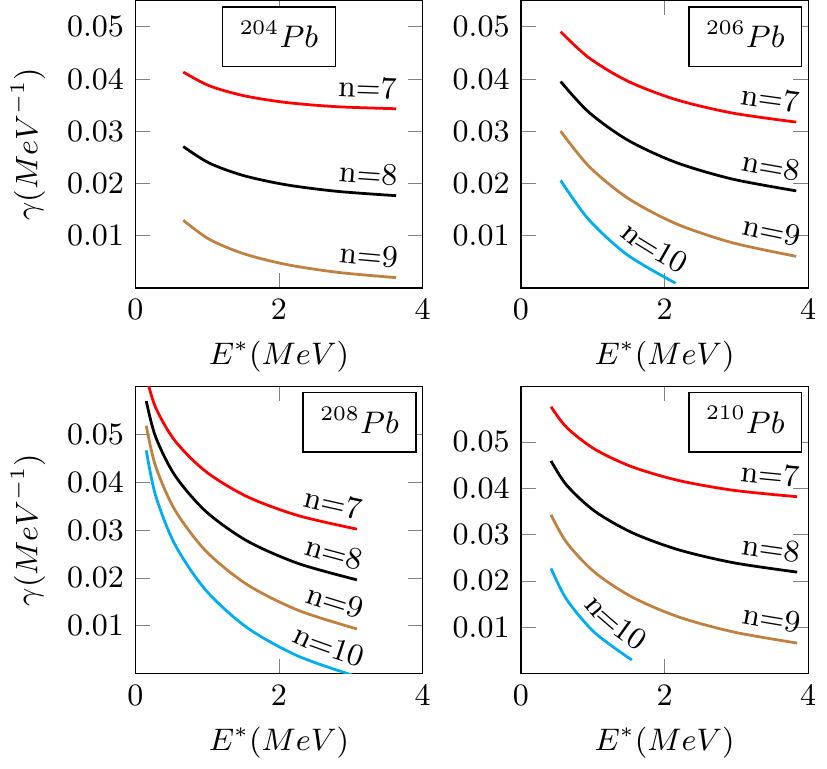}
\caption{Variation of $\gamma$ with excitation energy for isotopes of $Pb$ for different values of $n=A/\tilde{a}$. It can be seen that as excitation energy increases, the $\gamma$ values decrease, indicating the decrease of shell contributions. The shell contributions are lesser for higher $n$ values, suggesting that as a spherical symmetry is achieved, the damping approaches to zero.}
\label{fig:PbE}
\end{figure}

In $Sn$ isotopes (Fig. \ref{fig:Sn}), the $\delta W$ values of $^{112,114,116}Sn$ are closer to zero as compared to other nuclei. This clearly shows the contribution of the $\delta W$ values in the dependence of $\gamma$ with $n$. There is a steeper change of $\gamma$ with $n$ for these $Sn-$ isotopes.

In the case of $Pb$- isotopes (Fig. \ref{fig:Pb}), there is a decrease of $\gamma$ with the increase of $n$ values. These values come close to the predicted value of $\gamma\approx 0.05$ \cite{rout} for between $n=6$ to $n=7$ for $^{204-210}Pb$, and these values approach zero as $n$ approaches 10. For $^{182}Pb$, these values tend to zero much faster due to the $\delta W$ value being closer to zero.

Fig. \ref{fig:PbE} shows the variation of $\gamma$ with excitation energy $E^{*}$ for various values of $n$ for $Pb$ isotopes. Here it is visible that the least damping, and hence closer is the value of the level density parameter to its asymptotic value, is observed for $n=10$. These low values of $\gamma$ are achieved at $n=9$ for $^{204}Pb$. This damping also decreases with the increase in excitation energy, hinting the melting of shell effects with the increase in $E^*$.




\begin{figure}[h]
\centering

\includegraphics{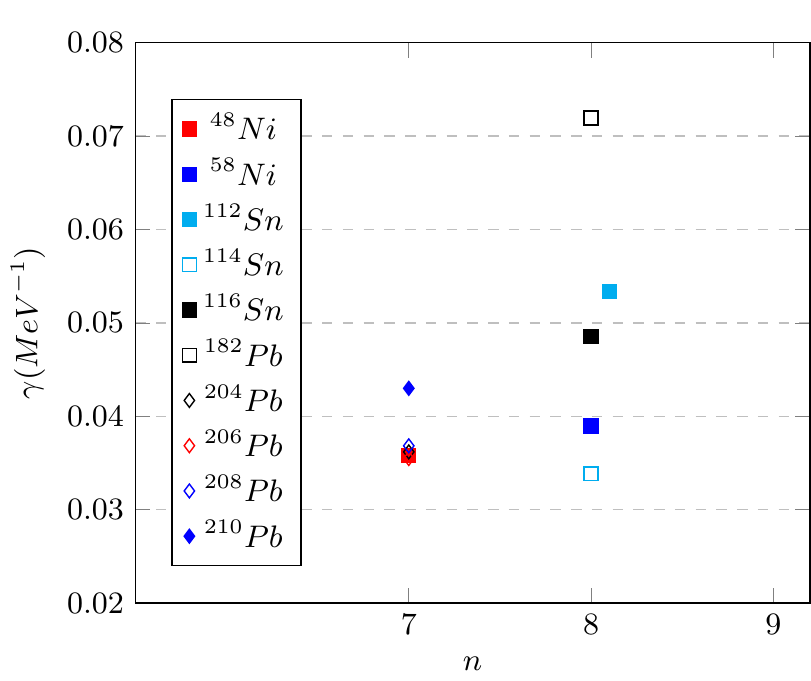}
\caption{The energy average values of $\gamma$ for the best $n$ values for the 10 nuclei.}
\label{fig:allav}
\end{figure}

The Gilbert Cameron Model predicts the parameter $\gamma$ by using the Fermi-gas model \cite{gil}. This model predicts that the values of $\gamma$ can be upto 0.1. We tune to find the best $n$-value for which the average $\gamma$ value over energies 0-10 $MeV$ best lie in this suggested range. 

Fig. \ref{fig:allav} shows the the average values of $\gamma$ of different nuclei for their best $n$ value. All the nuclei show their best values at $n=7,8$. The $Pb$- isotopes, except for $^{182}Pb$, lie on $n=7$ along with $^{48}Ni$. All considered $Sn$- isotopes along with $^{58}Ni$ and $^{182}Pb$, lie on $n=8$.

\begin{figure}[h]
\centering

\includegraphics{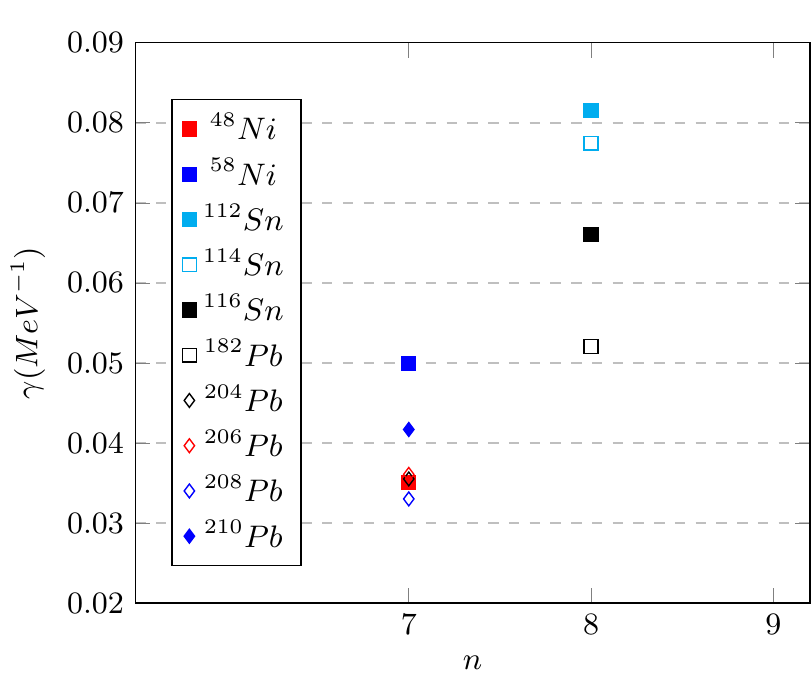}
\caption{The values of $\gamma$ for the best $n$ values for 10 nuclei at around 2 $MeV$ excitation energy.}
\label{fig:all2}
\end{figure}

 Next, we fix the excitation energy $E^{*}=2~ MeV$ and then find the best value of $n$ for which a specific nucleus has $\gamma$ in the region of experimentally extracted values. These best values for excitation energy $2~MeV$ are plotted in Fig. \ref{fig:all2}. As per our calculations, best $n$ differs from 10 for $Ni$-, $Sn$- and $Pb$- isotopes, which implies the role of pairing correlations and deformation must be included in the potential as well to have better agreement with the experimental values.

\section{Asymptotic behaviour of $n$}
The asymptotic value of the level density parameter, $\tilde{a}$ is given by $\tilde{a}=A/n$. At high excitation energies, since the shell effects melt away, the contribution of the $\gamma$-term becomes zero. The semiclassical formalism gives us excitation energy dependence of the level density and this level density parameter at high energies ($E^{*}>10~MeV$) correspond to the melted shell regime where the level density parameter approaches $\tilde{a}$. For higher energy dependence of $n$, we use the relation of $\tilde{a}$, and find the dependence of $n$ with excitation energy. Such systematics give hints of the form of the nuclear potential corresponding to harmonic oscillator ($n=10$) or of Woods-Saxon type ($n=8$). In deformed nuclei, such studies have been useful to understand shape transitions of nuclei with excitation energy \cite{banerjee}.

\begin{figure}[ht!]
\centering

\includegraphics{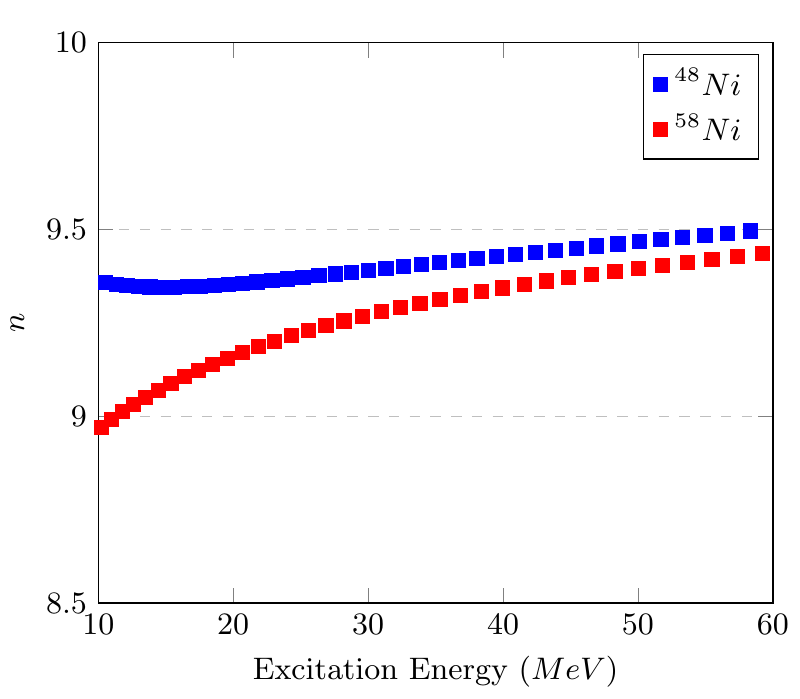}
\caption{Variation $n$ for $Ni-$isotopes with $E^*$ at high excitation energies. $^{48,58}Ni$ show values approaching closer to $n=10$, which corresponds to oscillator potential.}
\label{fig:Ni_n}
\end{figure}

\begin{figure}[ht!]
\centering

\includegraphics{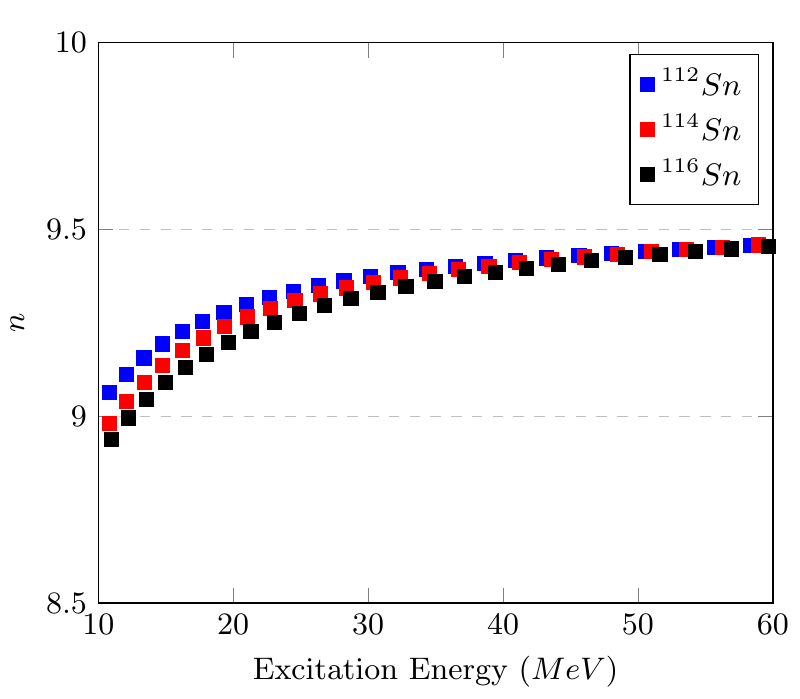}
\caption{Variation of $n$ for $Sn-$isotopes with $E^*$ at high excitation energies. $^{112,114,116}Sn$ show values approaching closer to $n=10$, which corresponds to a harmonic oscillator potential.}
\label{fig:Sn_n}
\end{figure}

\begin{figure}[ht!]
\centering
\includegraphics{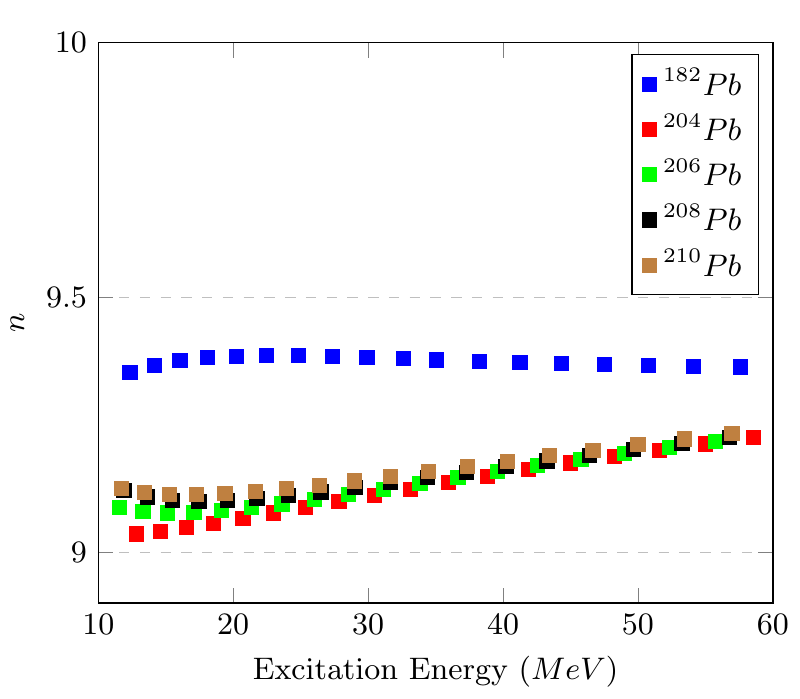}
\caption{Variation $n$ for $Pb-$isotopes with $E^*$ at high excitation energies. $Pb$- isotopes show the asymptotic values between $n=9$ and $n=9.5$, suggesting a potential between harmonic oscillator and Woods-Saxon. $^{182}Pb$ shows more closeness to the oscillator potential.}
\label{fig:Pb_n}
\end{figure}

From Figures \ref{fig:Ni_n} ,\ref{fig:Sn_n} and \ref{fig:Pb_n}, we see that the asymptotic values of level densities generally approach a constant value with increase in excitation energy, hinting a similar nuclear potential guiding them throughout the excitation energy. They remain between 9 and 10 for $^{48,58}Ni$,  $^{112,114,116}Sn$ and $^{182-210}Pb$.

A calculation on the same lines for a known deformed nucleus, like $^{56}Ni$, which is known to exhibit shape coexistence \cite{taka}, shows asymptotic level density follows $n=12$ trend. This may hint towards the contribution of pairing and deformation in the nuclear potential.

\section{Conclusions}
We have calculated the damping coefficient, $\gamma$ appearing in the Ignatyuk fitting prescription for ten nuclei involving $Ni$-, $Sn$- and $Pb$- isotopes. These values of the damping parameter are found in agreement with values used by experimentalists while analyzing the nuclear data in case of $Pb$- isotopes.  Further, the excitation energy dependence of $\gamma$ has been studied. As the excitation energy increases, the shell effects start to vanish, which means that the level density parameter should approach the value of $\tilde{a}$. This means that as excitation energy increases, $\gamma$ should decrease and tend to zero. This is demonstrated for many nuclei, an important illustrative instance is shown for the $Pb$- isotopes. As the value of $n$ approaches $n=10$, it is seen that the $\gamma$ values attains the minimum values, suggesting that the nucleus retains its spherical symmetry. We have further studied the asymptotic behaviour of $n$, which shows the variation of $n$ with excitation energy. To reiterate, the formula given here is based on a well-known trace formula \cite{brack-jain, brack}, which has no adjustable parameter. 

\section{Acknowledgements}
NRD acknowledges gratefully the funding for research from the Department of Atomic Energy and University of Mumbai - Bhabha Atomic Research Centre collaboration.


\begin{thebibliography}{10}

\bibitem{bohr}
Niels Bohr.
\newblock Neutron capture and nuclear constitution.
\newblock {\em Nature}, 137:344--348, 1936.

\bibitem{lang}
JMB Lang and KJ~Le~Couteur.
\newblock Statistics of nuclear levels.
\newblock {\em Proceedings of the Physical Society. Section A}, 67(7):586,
  1954.

\bibitem{zelevinsky}
V~Zelevinsky and S~Karampagia.
\newblock Nuclear level density, quantum chaos and related physics.
\newblock In {\em Journal of Physics: Conference Series}, volume 966, page
  012032. IOP Publishing, 2018.

\bibitem{bethe}
HA~Bethe.
\newblock An attempt to calculate the number of energy levels of a heavy
  nucleus.
\newblock {\em Physical Review}, 50(4):332, 1936.

\bibitem{eric}
Torleif Ericson.
\newblock A statistical analysis of excited nuclear states.
\newblock {\em Nuclear Physics}, 11:481--491, 1959.

\bibitem{al}
SI~Al-Quraishi, SM~Grimes, TN~Massey, and DA~Resler.
\newblock Level densities for 20<\~{} a<\~{} 110.
\newblock {\em Physical Review C}, 67(1):015803, 2003.

\bibitem{Audi}
Georges Audi, AH~Wapstra, and C~Thibault.
\newblock The ame2003 atomic mass evaluation:(ii). tables, graphs and
  references.
\newblock {\em Nuclear physics A}, 729(1):337--676, 2003.

\bibitem{Gutt}
M~Guttormsen, M~Hjorth-Jensen, E~Melby, J~Rekstad, A~Schiller, and S~Siem.
\newblock Entropy of thermally excited particles in nuclei.
\newblock {\em Physical Review C}, 63(4):044301, 2001.

\bibitem{gil}
A~Gilbert and AGW Cameron.
\newblock A composite nuclear-level density formula with shell corrections.
\newblock {\em Canadian Journal of Physics}, 43(8):1446--1496, 1965.

\bibitem{kawano}
Toshihiko Kawano, Satoshi Chiba, and Hiroyuki Koura.
\newblock Phenomenological nuclear level densities using the ktuy05 nuclear
  mass formula for applications off-stability.
\newblock {\em Journal of nuclear science and technology}, 43(1):1--8, 2006.

\bibitem{roy1}
M~Gohil, Pratap Roy, K~Banerjee, C~Bhattacharya, S~Kundu, TK~Rana, TK~Ghosh,
  G~Mukherjee, R~Pandey, H~Pai, et~al.
\newblock Angular momentum dependence of the nuclear level density in the a≈
  170--200 region.
\newblock {\em Physical Review C}, 91(1):014609, 2015.

\bibitem{roy2}
Pratap Roy, K~Banerjee, M~Gohil, C~Bhattacharya, S~Kundu, TK~Rana, TK~Ghosh,
  G~Mukherjee, R~Pandey, H~Pai, et~al.
\newblock Effect of collectivity on the nuclear level density.
\newblock {\em Physical Review C}, 88(3):031601, 2013.

\bibitem{Ignatyuk}
AV~Ignatyuk, GN~Smirenkin, and AS~Tishin.
\newblock Phenomenological description of energy dependence of the level
  density parameter.
\newblock {\em Yadernaya Fizika}, 21(3):485--490, 1975.

\bibitem{kataria}
SK~Kataria, VS~Ramamurthy, and SS~Kapoor.
\newblock Semiempirical nuclear level density formula with shell effects.
\newblock {\em Physical Review C}, 18(1):549, 1978.

\bibitem{vonach}
H~Vonach, M~Uhl, B~Strohmaier, BW~Smith, EG~Bilpuch, and GE~Mitchell.
\newblock Comparison of average s-wave resonance spacings from proton and
  neutron resonances.
\newblock {\em Physical Review C}, 38(6):2541, 1988.

\bibitem{nndc}
www.nndc.bnl.gov.

\bibitem{rout}
PC~Rout, DR~Chakrabarty, VM~Datar, Suresh Kumar, ET~Mirgule, A~Mitra, V~Nanal,
  SP~Behera, and V~Singh.
\newblock Measurement of the damping of the nuclear shell effect in the doubly
  magic pb 208 region.
\newblock {\em Physical review letters}, 110(6):062501, 2013.

\bibitem{Bohr-Mottelson2}
Aage Bohr and Ben~R Mottelson.
\newblock {\em Nuclear structure}, volume~2.
\newblock World Scientific, 1975.

\bibitem{jj}
Sudhir~R Jain, Ashok~K Jain, and Zafar Ahmed.
\newblock Nonlinear dynamics of high-j cranking model: A semi-classical
  approach.
\newblock {\em Physics Letters B}, 370(1-2):1--4, 1996.

\bibitem{deuteron}
Sudhir~R Jain.
\newblock Semiclassical deuteron.
\newblock {\em Journal of Physics G: Nuclear and Particle Physics}, 30(2):157,
  2004.

\bibitem{triton}
Nishchal~R Dwivedi, Harjeet Kaur, and Sudhir~R Jain.
\newblock Semiclassical triton.
\newblock {\em The European Physical Journal A}, 54(3):49, 2018.

\bibitem{zel}
Vladimir Zelevinsky, B~Alex Brown, Njema Frazier, and Mihai Horoi.
\newblock The nuclear shell model as a testing ground for many-body quantum
  chaos.
\newblock {\em Physics reports}, 276(2-3):85--176, 1996.

\bibitem{jp}
Sudhir~R Jain and Arun~K Pati.
\newblock Adiabatic geometric phases and response functions.
\newblock {\em Physical review letters}, 80(4):650, 1998.

\bibitem{npa}
Sudhir~R Jain.
\newblock Dissipation in finite fermi systems.
\newblock {\em Nuclear Physics A}, 673(1-4):423--451, 2000.

\bibitem{gulminelli}
Fran{\c{c}}ois Aymard, Francesca Gulminelli, and J{\'e}r{\^o}me Margueron.
\newblock In-medium nuclear cluster energies within the extended thomas-fermi
  approach.
\newblock {\em Physical Review C}, 89(6):065807, 2014.

\bibitem{suraud}
Eric Saraud, Peter Schuck, and Rainer~W Hasse.
\newblock On the temperature dependence of the level density parameter.
\newblock {\em Physics Letters B}, 164(4-6):212--216, 1985.

\bibitem{kaur}
Harjeet Kaur and Sudhir~R Jain.
\newblock Semiclassical theory of melting of shell effects in nuclei with
  temperature.
\newblock {\em Journal of Physics G: Nuclear and Particle Physics},
  42(11):115103, 2015.

\bibitem{brack-jain}
Matthias Brack and Sudhir~R Jain.
\newblock Analytical tests of gutzwiller’s trace formula for
  harmonic-oscillator potentials.
\newblock {\em Physical Review A}, 51(5):3462, 1995.

\bibitem{Bohr-Mottelson1}
Aage Bohr and Ben~R Mottelson.
\newblock {\em Nuclear structure}, volume~1.
\newblock World Scientific, 1969.

\bibitem{brack}
M~Brack and RK~Bhaduri.
\newblock {\em Semiclassical Physics (Boulder}.
\newblock Westview Press, 2003.

\bibitem{amann}
Ch~Amann and Matthias Brack.
\newblock Semiclassical trace formulae for systems with spin--orbit
  interactions: successes and limitations of present approaches.
\newblock {\em Journal of Physics A: Mathematical and General}, 35(29):6009,
  2002.

\bibitem{rag}
Ingemar Ragnarsson and Sven~Gvsta Nilsson.
\newblock {\em Shapes and shells in nuclear structure}.
\newblock Cambridge university press, 2005.

\bibitem{akjain}
A.~K. Jain, R.~K. Sheline, P.~C. Sood, and Kiran Jain.
\newblock Intrinsic states of deformed odd-a nuclei in the mass regions (151≤
  a≤ 193) and (a≥ 221).
\newblock {\em Reviews of Modern Physics}, 62(2):393, 1990.

\bibitem{banerjee}
K~Banerjee, Pratap Roy, Deepak Pandit, Jhilam Sadhukhan, S~Bhattacharya,
  C~Bhattacharya, G~Mukherjee, TK~Ghosh, S~Kundu, A~Sen, et~al.
\newblock Direct evidence of fadeout of collective enhancement in nuclear level
  density.
\newblock {\em Physics Letters B}, 772:105--109, 2017.

\bibitem{taka}
Takahiro Mizusaki, Takaharu Otsuka, Yutaka Utsuno, Michio Honma, and Takashi
  Sebe.
\newblock Shape coexistence in doubly-magic 56 ni by the monte carlo shell
  model.
\newblock {\em Physical Review C}, 59(4):R1846, 1999.

\end{thebibliography}
\end{document}